\documentclass[sigconf]{acmart}

\usepackage{adjustbox}
\usepackage{tikz}
\usetikzlibrary{shapes.geometric, arrows.meta, positioning}
\usepackage{multirow}
\usepackage{subcaption}
\usepackage{hyperref}
\acmConference[]{}{}

\title{Diabetic Retinopathy Detection Using CNN with Residual Block and DCGAN}
\author{Debjany Ghosh Aronno}
\email{debjanyaronno@gmail.com}
\affiliation{%
  \institution{Bangladesh University of Engineering \& Technology}
  \city{Dhaka}
  \country{Bangladesh}}
\authornote{Both authors contributed equally to this research.}
\author{Sumaiya Saeha}
\authornotemark[1]
\email{sumaiyasaeha@gmail.com}
\affiliation{%
  \institution{Bangladesh University of Engineering \& Technology}
  \city{Dhaka}
  \country{Bangladesh}}

\begin{document}
\begin{abstract}
    Diabetic Retinopathy (DR) is a major cause of blindness worldwide, caused by damage to the blood vessels in the retina due to diabetes. Early detection and classification of DR are crucial for timely intervention and preventing vision loss. This work proposes an automated system for DR detection using Convolutional Neural Networks (CNNs) with a residual block architecture, which enhances feature extraction and model performance. To further improve the model's robustness, we incorporate advanced data augmentation techniques, specifically leveraging a Deep Convolutional Generative Adversarial Network (DCGAN) for generating diverse retinal images. This approach increases the variability of training data, making the model more generalizable and capable of handling real-world variations in retinal images. The system is designed to classify retinal images into five distinct categories, from No DR to Proliferative DR, providing an efficient and scalable solution for early diagnosis and monitoring of DR progression. The proposed model aims to support healthcare professionals in large-scale DR screening, especially in resource-constrained settings.
    \end{abstract}
\maketitle

\section{Introduction}

Diabetic Retinopathy (DR) is a microvascular complication that occurs as a result of diabetes and is recognized as one of the leading causes of blindness worldwide. The disease primarily affects the blood vessels in the retina, leading to a range of vision problems. Early diagnosis and timely intervention are crucial in preventing irreversible vision loss, making DR detection a critical aspect of diabetic care. As DR progresses, it manifests as a spectrum of stages, starting from no DR, which indicates the absence of visible abnormalities, to mild, moderate, and severe stages, and finally to proliferative DR, the most advanced form where new, abnormal blood vessels start to grow in the retina. 

These different stages of DR are associated with various retinal abnormalities, such as microaneurysms, hemorrhages, exudates, and neovascularization, which require careful and precise examination. Accurate and automated detection of DR severity can significantly improve the efficiency and effectiveness of large-scale screening programs, particularly in resource-constrained environments where access to trained ophthalmologists and diagnostic equipment is limited. Automated systems can aid in early detection and help prioritize cases that require urgent intervention, thus reducing the burden on healthcare professionals and ensuring that at-risk patients receive the appropriate care in a timely manner.

In this work, we focus on the classification of retinal images into five distinct categories, ranging from No DR to Proliferative DR. These categories represent the varying degrees of severity and retinal changes associated with DR. By developing an automated system capable of accurately classifying retinal images into these five stages, we aim to enhance the ability to detect DR early and to monitor its progression over time, ultimately contributing to more effective management of diabetic patients and reducing the risk of blindness due to DR.
\subsection{Problem Definition}

Identifying and classifying DR severity levels from retinal fundus images is a challenging multiclass classification task. It requires distinguishing subtle visual patterns indicative of various DR stages, which may not be apparent to untrained eyes. Unlike traditional feature extraction methods, which rely heavily on manual input and domain expertise, end-to-end deep learning approaches like Convolutional Neural Networks (CNNs) directly learn discriminative features from data. CNN-based architectures are particularly well-suited for this problem due to their ability to capture spatial hierarchies in images.

\subsection{Literature Review and Gap Analysis}
The methods used to determine the severity of DR from fundus images can be broadly classified into two categories: 
\begin{itemize} 
\item Classification after component extraction 
\item Classification without component extraction \end{itemize} 
The first category focuses on detection by identifying various anatomical components within a fundus image. On the other hand, the second category involves direct detection from raw fundus images without the explicit identification of any components.  In the work of Ahmad et al. \cite{ahmad2019automatic} the severity of DR is detected from different properties of the anatomical components present in a fundus image following the first approach. In contrast, several studies have proposed CNN-based models for DR detection and classification following the second approach. For instance, Gulshan et al. \cite{gulshan2016development} introduced a CNN-based algorithm for DR detection with promising performance on large datasets. Similarly, Voets et al. \cite{voets2018replication} analyzed the reproducibility of CNNs in DR detection, while Lam et al. \cite{lam2018automated} demonstrated the effectiveness of transfer learning for DR classification. Papon et al. \cite{papon2019design}  developed a novel deep learning based DR severity detection model trained
 using their own loss function with a large heterogeneous dataset. These works often employ standard data augmentation techniques, such as rotation, flipping, and shearing, to address the inherent data imbalance in DR datasets. In the context of the future scope of the work of Kommaraju et al. \cite{kommaraju2024diabetic}, we aim to explore the use of residual blocks to enhance feature extraction and improve the model's ability to handle complex and imbalanced DR datasets, with the potential to improve classification accuracy and generalization.

However, a major challenge remains: DR datasets are typically highly imbalanced, with the majority of images belonging to the No DR category. Such imbalances can lead to model biases and suboptimal performance on underrepresented classes, such as Severe and Proliferative DR. While conventional augmentation techniques can help expand the dataset, they often fail to introduce sufficient diversity to capture the complexity of rare classes. Advanced data augmentation methods, such as those based on Generative Adversarial Networks (GANs) and Variational Autoencoders (VAEs), remain underexplored in DR classification. Few works have leveraged these techniques for creating synthetic training samples, which can enhance model robustness and address class imbalance more effectively. This gap highlights the need for novel approaches that integrate advanced augmentation strategies into DR detection pipelines.





\section{Materials and Methods}

\subsection{Dataset}

\begin{figure}[h!]
    \centering
    \resizebox{0.4\textwidth}{!}{\includegraphics{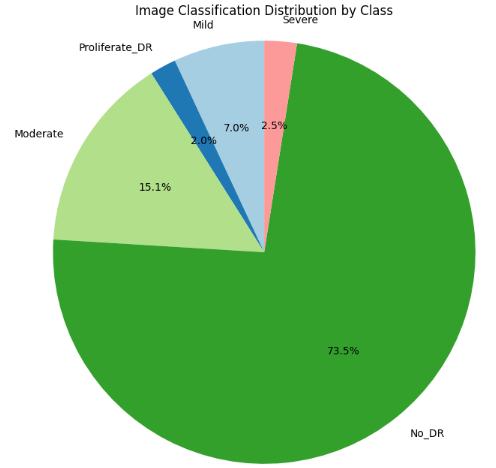}}
    \caption{Image classification distribution by class}
    \label{fig:pie}
\end{figure}

The dataset used for this study comprises retinal fundus images classified into five categories: No DR, Mild, Moderate, Severe, and Proliferative DR. The dataset includes 35126 images, with the following distribution of classes: 25810 No DR images, 2443 Mild images, 5292 Moderate images, 873 Severe images, and 708 Proliferative DR images. The image classification distribution is shown in Fig~\ref{fig:pie}. The images were collected from \href{https://www.kaggle.com/datasets/sovitrath/diabetic-retinopathy-2015-data-colored-resized}{\color{blue}this dataset}. The dataset was split into training, validation, and testing subsets using a 80:10:10 split. Each image was resized to a fixed dimension of 224x224 to ensure consistency during model training.

\subsection{Preprocessing}

Medical images are often challenging to analyze due to their complexity and the inherent difficulties in processing them. Therefore, preprocessing techniques are essential for improving the features of the images for classification and ensuring consistency across the dataset.

In the proposed model, several preprocessing techniques are employed: (1) \textbf{Circle-Crop}, (2) \textbf{Median Subtraction}, (3) \textbf{Gamma Correction}, and (4) \textbf{Adaptive Histogram Equalization}. The circle-crop technique helps remove unnecessary background noise and standardizes all images to a uniform size. Median Subtraction uses a median filter to reduce noise without affecting the edges of the image, providing a faster alternative to other filters. Gamma Correction is applied to adjust the pixel saturation in non-linear ways, controlling the relationship between the pixel values and the actual brightness of the image. Finally, Adaptive Histogram Equalization improves the contrast of the image by considering multiple histograms to enhance local details.

\subsection{Data Augmentation}

To address the significant class imbalance in the dataset, advanced data augmentation techniques were employed. While conventional augmentations such as rotation, flipping, shearing, zooming, width shift, height shift and brightness adjustments were applied to all classes, the underrepresented classes (all except No DR) were further augmented using Deep Convolutional GANs (DC-GANs). These GAN-generated synthetic images enhanced the diversity of the training set, helping the model learn more robust representations for rare classes. The architecture consisted of a generator and a discriminator trained adversarially to create high-quality, realistic retinal fundus images. 

\begin{itemize}
    \item \textbf{Generator:} The generator consisted of transpose convolutional layers, each followed by batch normalization and ReLU activations. The final layer used a \texttt{tanh} activation to output synthetic images with pixel values normalized between -1 and 1.
    \item \textbf{Discriminator:} The discriminator employed convolutional layers with Leaky ReLU activations and dropout layers to enhance robustness. The final layer used a sigmoid activation to classify images as real or synthetic.
\end{itemize}

\textbf{Training Details:}
\begin{itemize}
    \item \textbf{Number of Epochs:} 10 epochs with 3750 steps per epoch were used to ensure the generator learned to produce high-quality images. 
    \item \textbf{Batch Size:} A batch size of 4 was chosen for both the generator and discriminator training.
    \item \textbf{Learning Rate:} The Adam optimizer was used with a learning rate of 0.0002 and a $\beta_1$ value of 0.5 for stable training.
    \item \textbf{Latent Vector Size:} The generator was fed a 100-dimensional random latent vector sampled from a uniform distribution.
    \item \textbf{Training Images:} Images from underrepresented classes were resized to $128 \times 128$ pixels for GAN training.

\end{itemize}

\begin{figure}[h!]
    \centering
    \begin{minipage}{0.45\textwidth}
        \centering
        \resizebox{0.45\textwidth}{!}{\includegraphics{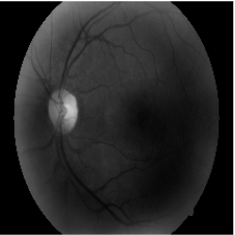}}
        \caption{Real image}
        \label{fig:real}
    \end{minipage}\hfill
    \begin{minipage}{0.45\textwidth}
        \centering
        \resizebox{0.45\textwidth}{!}{\includegraphics{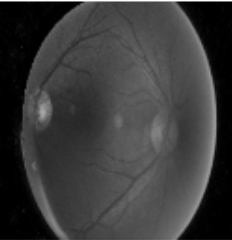}}
        \caption{Generated fake image}
        \label{fig:fake}
    \end{minipage}
\end{figure}

\begin{figure}[h!]
    \centering
    \resizebox{0.4\textwidth}{!}{\includegraphics{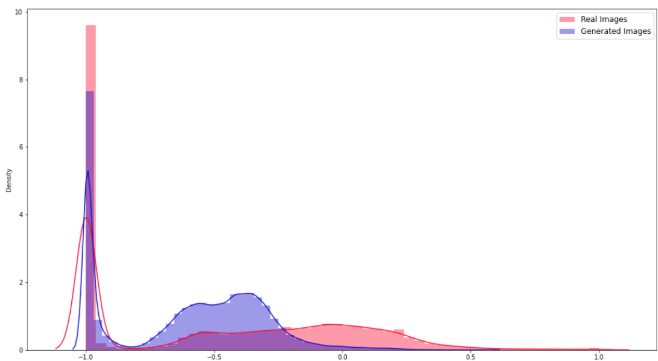}}
    \caption{Image classification distribution by class}
    \label{fig:pid}
\end{figure}

The GAN training process involved alternating updates to the generator and discriminator, with the discriminator being trained to distinguish between real and synthetic images, and the generator being trained to produce images that could fool the discriminator. After training, the synthetic images were added to the training set for underrepresented classes, significantly enhancing class balance and improving model performance. The example of a real and fake image is shown in Fig~\ref{fig:real} and Fig~\ref{fig:fake}. As shown in Fig~\ref{fig:pid}, the pixel intensity distributions of real and generated images were also quite similar, indicating the effectiveness of the GAN in capturing the underlying data distribution. The x-axis denotes the pixel intensity values, ranging from -1.0 to 1.0. The y-axis shows density of pixels with a given intensity value. Red line and histogram represent the pixel intensity distribution of real images, whereas blue line and histogram represent the pixel intensity distribution of images generated by the DC-GAN. The image suggests that the DC-GAN is able to generate images with a similar overall intensity distribution to real images. However, there are some differences in peak intensity and distribution width, which may indicate potential limitations of the model.
\subsection{Proposed Architecture}

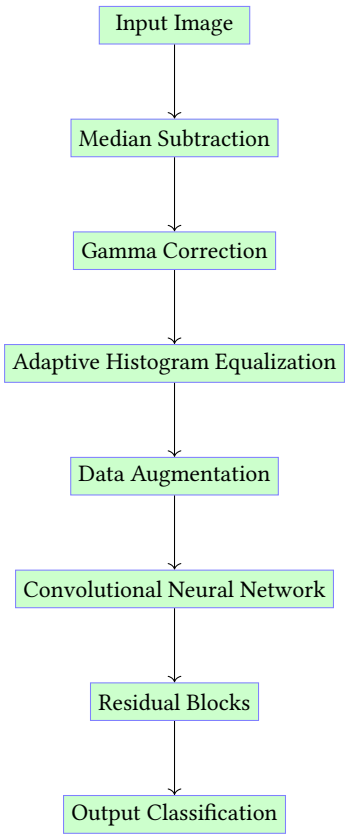
\begin{figure}[ht]
    \centering
    \begin{tikzpicture}[node distance=1.5cm, every node/.style={rectangle, draw=blue!50, fill=green!20, minimum width=2cm, minimum height=0.5cm},
    every edge/.style={draw=blue!50, ->, shorten <= 0.1cm, shorten >= 0.1cm}]
        
        \node (input) {Input Image};
        \node (median) [below of=input] {Median Subtraction};
        \node (gamma) [below of=median] {Gamma Correction};
        \node (ahe) [below of=gamma] {Adaptive Histogram Equalization};
        \node (aug) [below of=ahe] {Data Augmentation};
        \node (cnn) [below of=aug] {Convolutional Neural Network};
        \node (res) [below of=cnn] {Residual Blocks};
        \node (output) [below of=res] {Output Classification};

        \draw[->] (input) -- (median);
        \draw[->] (median) -- (gamma);
        \draw[->] (gamma) -- (ahe);
        \draw[->] (ahe) -- (aug);
        \draw[->] (aug) -- (cnn);
        \draw[->] (cnn) -- (res);
        \draw[->] (res) -- (output);

    \end{tikzpicture}
    \caption{Architecture for Image Classification}
    \label{fig:archi}
\end{figure}

The proposed model is a Convolutional Neural Network (CNN) enhanced with residual blocks. The architecture is designed to capture fine-grained features in retinal fundus images, enabling effective discrimination between the five DR severity levels. The network consists of the following components:
\begin{itemize}
    \item \textbf{Input Layer:} Accepts resized retinal images of dimension 224x224x1.
    \item \textbf{Convolutional Layers:} Employs techniques like zero-padding and max pooling to preserve information and reduce dimensionality. It utilizes activation functions like ReLU to introduce non-linearity.
    \item \textbf{Residual Blocks:} Added to mitigate the vanishing gradient problem. Each block includes a convolutional block and two identity blocks. It helps in training deeper networks effectively.
    \item \textbf{Identity Blocks:} It is similar to residual blocks but without the shortcut connection, maintains the dimensionality of the input.
    \item \textbf{Fully Connected Layers:} It transforms the high-dimensional feature maps into a lower-dimensional representation and prepares the data for the final classification layer.
    \item \textbf{Output Layer:} Employs a softmax layer to output probabilities for each of the five classes (No DR, Mild, Moderate, Severe, Proliferative DR).
\end{itemize}
The model's architecture is illustrated in Fig~\ref{fig:archi}.

\subsection{Training Details}

The model was implemented using the TensorFlow framework with Keras as the high-level API. The training process was conducted on Kaggle's cloud environment equipped with dual NVIDIA Tesla T4 GPUs, each with 16 GB of VRAM. This setup provided sufficient computational power to handle the intensive operations required for training the deep learning model efficiently.

The training process involved the following steps:
\begin{itemize}
    \item \textbf{Loss Function:} Categorical Cross-Entropy was used as the loss function, suitable for multi-class classification problems. This function penalizes the model for incorrect predictions, ensuring that the predicted probability distributions align closely with the actual labels.
    
    \item \textbf{Optimizer:} The Adam optimizer was utilized with an initial learning rate of 0.001. A learning rate scheduler was employed to reduce the learning rate by a factor of 0.1 every 10 epochs, helping the model converge to a global minimum.
    
    \item \textbf{Batch Size:} A batch size of 32 was selected after hyperparameter tuning. This batch size provided a balance between computational efficiency and memory usage on the GPU.
    
    \item \textbf{Number of Epochs:} The model was trained for 50 epochs. Early stopping with a patience of 15 epochs was used to terminate training when the validation loss did not improve, thereby avoiding overfitting. 
    
    \item \textbf{Data Augmentation:} 
        \begin{itemize}
            \item Standard augmentations such as rotation (up to 20 degrees), width and height shifts (up to 20\%), shear transformations, zoom (up to 20\%), and horizontal flipping were applied to diversify the training dataset.
            \item For under-represented classes, advanced augmentation using DC-GANs was employed to synthetically generate realistic samples.
        \end{itemize}
    
    \item \textbf{Evaluation Metrics:} To comprehensively evaluate model performance, the following metrics were calculated for each class:
        \begin{itemize}
            \item \textit{Accuracy:} The proportion of correctly classified samples.
            \item \textit{Precision:} The ratio of true positives to the sum of true positives and false positives, indicating the reliability of positive predictions.
            \item \textit{Recall:} The ratio of true positives to the sum of true positives and false negatives, highlighting the model's sensitivity.
            \item \textit{F1-Score:} The harmonic mean of precision and recall, providing a balanced measure for imbalanced datasets.
        \end{itemize}
    
    \item \textbf{Hardware and Software:} The training was performed on two NVIDIA T4 GPUs using TensorFlow 2.9.1. The code was executed on a system running Kaggle's environment with Python 3.9.
\end{itemize}
\begin{figure}[h!]
    \centering
    \resizebox{0.5\textwidth}{!}{\includegraphics{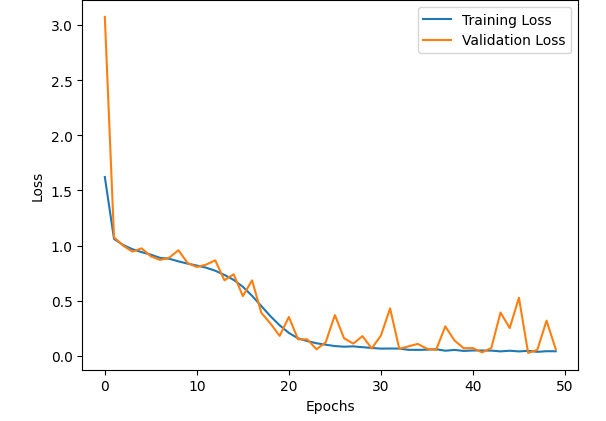}}
    \caption{Training and Validation Loss Curves}
    \label{fig:train_vs_val}
\end{figure}
\subsection{Evaluation and Validation}

To ensure robust performance, the model was validated on the separate validation set during training. Early stopping was applied to prevent overfitting, with the patience parameter set to 15. As shown in Fig~\ref{fig:train_vs_val}, the training loss decreased steadily, while the validation loss stabilized after around 20 epochs, indicating optimal model performance. The final model was evaluated on the test set, and a confusion matrix was generated to analyze misclassifications across the five classes. 

\section{Results and Discussion}

\subsection{Metrics Considered for Performance Evaluation}
The performance of the proposed model is evaluated using various metrics, including accuracy, recall, F1-score, and precision. The definitions of these metrics are as follows:

\begin{itemize}
    \item \textbf{Accuracy:} The ratio of correctly predicted instances to the total instances. It is given by:
    \[
    \text{Accuracy} = \frac{TP + TN}{TP + TN + FP + FN}
    \]
    where $TP$, $TN$, $FP$, and $FN$ represent the number of true positives, true negatives, false positives, and false negatives, respectively.
    
    \item \textbf{Recall:} The ratio of correctly predicted positive instances to all actual positive instances. It is given by:
    \[
    \text{Recall} = \frac{TP}{TP + FN}
    \]
    
    \item \textbf{Precision:} The ratio of correctly predicted positive instances to the total predicted positive instances. It is given by:
    \[
    \text{Precision} = \frac{TP}{TP + FP}
    \]

    \item \textbf{F1-score:} The harmonic mean of precision and recall, providing a balanced measure. It is given by:
    \[
    \text{F1-score} = 2 \times \frac{\text{Precision} \times \text{Recall}}{\text{Precision} + \text{Recall}}
    \]
\end{itemize}

\subsection{Performance Evaluation}
The performance metrics, including F1-score, recall, and precision, for the proposed model are presented in Table~\ref{tab:class-rep}. The confusion matrix is shown in Table~\ref{tab:confusion-matrix}.

\begin{table}[h]
\centering
\caption{Classification Report}
\label{tab:class-rep}
\begin{tabular}{l|cccc}
\hline
Class & Precision & Recall & F1-score & Support \\
\hline
No DR & 0.998 & 0.999 & 0.998 & 2562 \\
Moderate & 0.978 & 0.95 & 0.964 & 526 \\
Mild & 0.94 & 0.97 & 0.955 & 263 \\
Proliferative DR & 0.88 & 0.96 & 0.918 & 75 \\
Severe & 0.954 & 0.954 & 0.954 & 87 \\
\hline
accuracy &  &  & 0.987 & 3513 \\
\hline
\end{tabular}
\end{table}

\begin{table}[h!]
\centering
\begin{adjustbox}{width=0.45\textwidth}
\begin{tabular}{|c|c|c|c|c|c|c|}
\hline
\multicolumn{2}{|c|}{\multirow{2}{*}{}} & \multicolumn{5}{c|}{Predicted labels} \\ \cline{3-7}
\multicolumn{2}{|c|}{} & \textbf{No DR} & \textbf{Moderate} & \textbf{Mild} & \textbf{Proliferative DR} & \textbf{Severe} \\ \hline
\multirow{5}{*}{True} & \textbf{No DR} & 2560 & 1 & 0 & 1 & 0 \\ \cline{2-7}
 & \textbf{Moderate} & 2 & 500 & 15 & 6 & 3 \\ \cline{2-7}
 & \textbf{Mild} & 1 & 5 & 255 & 1 & 1 \\ \cline{2-7}
 labels & \textbf{Proliferative DR} & 0 & 3 & 0 & 72 & 0 \\ \cline{2-7}
 & \textbf{Severe} & 1 & 2 & 0 & 1 & 83 \\ \hline
\end{tabular}
\end{adjustbox}
\caption{Confusion Matrix for Classification Results}
\label{tab:confusion-matrix}
\end{table}



\subsection{Discussion}
Overall, our model demonstrates strong performance in cases of No DR and Moderate, but its performance tends to fluctuate in other categories. However, the model could benefit from further optimization to handle the issues identified in the analysis. One of the key challenges we encountered was selecting a suitable GAN architecture; after careful consideration, we opted for DCGAN. Ensuring the realism of the GAN-generated images to effectively augment the training set proved to be demanding and required extensive visual inspection. Additionally, training GANs was both computationally intensive and time-consuming. While we initially experimented with generating augmented images using RGB inputs, this approach resulted in noisy outputs. Consequently, we focused on grayscale images, which yielded more reliable results.

\section{Conclusion}

In this study, we employed DCGAN to generate realistic images for augmenting the training dataset. After considering various GAN architectures, we faced challenges in ensuring the quality of the generated images, which required extensive visual inspection. While our model performed well in the No DR category without the need for GAN augmentation, we used GAN-generated images to improve performance in the other categories. Ultimately, we utilized the full dataset for model training.

Future work could focus on refining the GAN architecture to generate higher-quality images and reduce noise, especially when using RGB images. Further analysis and fine-tuning may be needed to improve the quality of the generated images. Expanding the dataset and leveraging more computational resources would help improve model performance and scalability. 

\bibliographystyle{ACM-Reference-Format}
\bibliography{reference}

\end{document}